# Code-Based Automated Program Fixing

Yu Pei · Yi Wei · Carlo A. Furia · Martin Nordio · Bertrand Meyer

August 2011

**Abstract**

Many programmers, when they encounter an error, would like to have the benefit of automatic fix suggestions—as long as they are, most of the time, adequate. Initial research in this direction has generally limited itself to specific areas, such as data structure classes with carefully designed interfaces, and relied on simple approaches.

To provide high-quality fix suggestions in a broad area of applicability, the present work relies on the presence of contracts in the code, and on the availability of static and dynamic analyses to gather evidence on the values taken by expressions derived from the code.

The ideas have been built into the AutoFix-E2 automatic fix generator. Applications of AutoFix-E2 to general-purpose software, such as a library to manipulate documents, show that the approach provides an improvement over previous techniques, in particular purely model-based approaches.

## 1 Introduction

Debugging—the activity of finding and correcting errors in programs—is so everyday in every programmer's job that any improvement at automating even parts of it has the potential for a significant impact on productivity and software quality.

While automation remains formidably difficult in general, the last few years have seen the first successful attempts at providing completely automated debugging in some situations. This has been achieved with the combination of several techniques developed independently: automated testing to detect errors, fault localization to locate instructions responsible for the errors, and dynamic analysis to choose suitable corrections among those applicable to the faulty instructions. Consider, for example, a routine (method) which removes the last element in a linked list by getting a reference and deallocating it. Random testing tries the routine on an empty list and exposes an error; fault localization suggests that the problem is deallocating the last element when it is void (null); dynamic analysis suggests to change the behavior of the routine so that deallocation is performed only when the last element exists.

A few premises make such automated debugging techniques work in practice. First, the majority of errors in programs admit simple fixes [6], consisting in adding or modifying one or two instructions; correspondingly, generating the set of possible "small" corrections exhaustively is often computationally feasible. Second, the availability of contracts (pre and postconditions, class invariants) can dramatically improve the accuracy of both error detection and fault localization.

Our previous work in this area [5, 21] takes advantage of these observations to perform an analysis of faults in object-oriented programs with contracts and correct them. The analysis constructs an abstract model of correct and incorrect executions, which summarizes the information about the program state at various locations in terms of state invariants. The invariants express the values returned by *public queries* (functions) of a class—the same functions used by developers in the contracts that document the implementation. The comparison of the invariants characterizing correct and incorrect runs suggests how to fix errors: whenever the state signals the "incorrect invariant", execute actions to avoid triggering the error. A behavioral model of the class, also relying on state invariants, suggests the applicable "recovery" actions. We call this approach to automated program fixing *model-based*, given that a model, based on state invariants, abstracts the correct and incorrect visible behavior. In the example of the linked list, assume that the class has a query *empty*, which returns true when the list contains no elements, and that the correct and incorrect runs respectively have invariants **not** *empty* and *empty*, because the failure occurs precisely when the list is empty. A reasonable fix consists in adding a conditional statement which guards the deallocation instruction and executes it only when **not** *empty* is the case.

The efficacy of model-based fixing fundamentally depends on the quality of the public interfaces, because invariants are mostly based on public queries. The present paper introduces a more general approach to automated fixing which works successfully even for classes with few public queries. The approach is still based on the dynamic analysis of correct and incorrect runs. However, rather than merely monitoring the value of queries, the analysis proactively gathers evidence in terms of values taken by *expressions* appearing in the program text. An algorithm built upon fault localization techniques—based on static and dynamic analysis—ranks expressions and their values according to their likelihood of being indicative of error. The expressions ranking highest are prime candidates to guide the generation of fixes: when an expression takes a "suspicious" value, execute actions that change the value to "unsuspicious". We call this novel approach *code-based* to designate the white-box search for information denoting faults in the program text. In the sketched example of the linked list, code-based techniques can build a fix even if a query *empty* is not available, by choosing to monitor the value of the expression denoting the reference to the last element in the list.

The designations "model-based" and "code-based" schematize the essential differences between the two approaches, but it is important to remark that the latter is essentially an *extension* (and improvement) of the former: code-based techniques also exploit information in the form of state invariants and public queries to reproduce the results of model-based techniques when these are successful.

We implemented code-based fixing in the tool AutoFix-E2, successor to AutoFix-E [21] which implemented model-based techniques. The experiments in Section 4 demonstrate that code-based techniques can automatically fix more errors than model-based approaches, even beyond data structure implementations—the natural target of model-based and random-testing techniques, for their rich public interfaces.

The paper is organized as follows: Section 2 introduces code-based techniques with an example of fix which is beyond the capabilities of model-based techniques; Section 3 details the ingredients of code-based fixing and how they are combined; Section 4 presents an experimental evaluation of the implementation AutoFix-E2; Section 5 discusses related work; Section 6 outlines future work.

## 2  Automated Fixing: an example

This section illustrates two faults fixed by AutoFix-E2; the example shows the edge of code-based techniques in specific fixing scenarios and is used throughout the paper.

### 2.1  Two Errors in a Routine

The EiffelBase class *TWO_WAY_SORTED_SET* implements a set data structure with a doubly-linked list. An internal cursor *index* (an integer attribute) is useful to navigate the content of the set: the actual elements occupy positions 1 to *count* (another integer attribute, storing the total number of elements in the set), whereas the indexes 0 and *count* + 1 correspond to the positions *before* the first element and *after* the last. Listing 1 shows the routine *move_item* of this class, which takes an argument $v$ of generic type $G$ that must be a reference to an element already stored in the set; the routine then moves $v$ from its current (unique) position in the set to the immediate left of the internal cursor *index*. For example, if the set is $\langle a, b, c, v \rangle$ and *index* is 2 upon invocation, *move_index* $(v)$ changes the set to $\langle a, v, b, c \rangle$. The routine's precondition (**require**) formalizes the constraint on the input. After saving the cursor position as the local variable *idx*, the loop in lines 7–10 performs a linear search for the element $v$ using the internal cursor: when the loop terminates, *index* denotes $v$'s position in the set. The three routine calls on lines 12–14 complete the work: *remove* takes $v$ out of the set; *go_i_th* restores *index* to its original value *idx*; *put_left* puts $v$ back in the set to the left of the position *index*.

AutoTest [17] reveals, completely automatically, two errors in this implementation of *move_item*. The first error is due to the fact that calling *remove* decrements the *count* of elements in the set by one. AutoTest produces a test (shown in Figure 1) that calls *move_item* when *index* equals *count* + 1; after $v$ is removed, this value is not a valid position because it exceeds the new value of *count* by two, while a valid cursor ranges between 0 and *count* + 1. The test violates *go_i_th*'s precondition (line 17), which enforces the consistency constraint on *index*, when invoking it on line 13.

The second error occurs when *index* has value 0, denoted by the boolean query *before* (line 19); this is a valid position for *go_i_th* but not for *put_left*, because there is no position "to the left of 0" where $v$ can be re-inserted: the call to *put_left* on line 14 violates its precondition (line 18).

### 2.2  Code-Based Fixing at Work

The fault revealed in the invocation of *go_i_th* is actually a special case of a more general error which occurs whenever $v$ appears in the set in a position to the left

Listing 1: Routines of *TWO_WAY_SORTED_SET*.

```
1   move_item (v: G)
2       −− Move 'v' to the left of cursor.
3       require v ≠ Void ; has (v)
4       local idx: INTEGER ; found: BOOLEAN
5       do
6           idx := index
7           from start until found or after loop
8               found := (v = item)
9               if not found then forth end
10          end
11          check found and not after end
12          remove
13          go_i_th (idx)
14          put_left (v)
15      end
16
17  go_i_th (i: INTEGER) require 0 ≤ i ≤ count + 1
18  put_left (v: G) require not before
19  before: BOOLEAN do Result := (index = 0) end
```

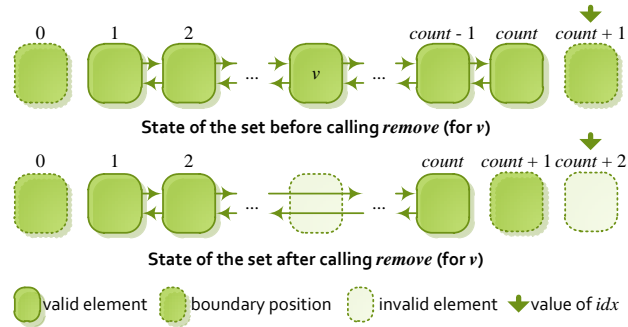

Figure 1: Calling *remove* in *move_item* when $index = count + 1$ holds initially makes the following invocation of *go_i_th* (*idx*) violate a precondition.

of the initial value of *index*: even if $index \leq count$ initially, *put_left* will insert *v* in the wrong position as a result of *remove* decrementing *count*—which indirectly shifts the index of every element after *index* to the left by one. For example, if *index* is 3 initially, $\langle a, v, b, c \rangle$ becomes $\langle a, b, v, c \rangle$, instead of staying unchanged, after calling *move_item* (*v*). Such states leading to erroneous behavior go undetected by AutoTest because the developers of *TWO_WAY_SORTED_SET* provided an incomplete postcondition; more generally, the class lacks a query to characterize the fault condition in general terms. Nonetheless, AutoFix-E2 can completely correct the error, beyond the specific case reported by the failed test: it builds the expression $idx > index$ to characterize the error state and generates the corresponding fix, introduced before line 13, which re-scales *idx* to reflect the fact that the object in position *idx* has been shifted left.

$$\textbf{if } idx > index \textbf{ then } idx := idx - 1 \textbf{ end}$$

The error in the invocation of *put_left*, on the other hand, is accurately characterized by the public query *before*, which returns **True** whenever the call on line 14 triggers a precondition violation. The correction suggested automatically by AutoFix-E2 adds the instruction **if** *before* **then** *forth* **end** right before

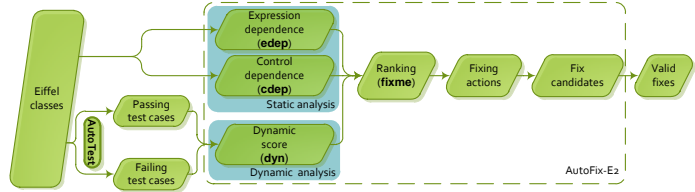

Figure 2: How code-based fixing works. Run AutoTest to automatically generate passing and failing test cases for the input Eiffel classes (Section 3.1); extract a set of *expressions* from the text of the classes' routines (Section 3.2); compute the *expression dependence* score edep between expressions—measuring their syntactic similarity (Section 3.3.2)—and the *control dependence* score cdep between program locations—measuring how close they are on the control-flow graph (Section 3.3.1); compute the *dynamic score* dyn between expressions, measuring how more often an expression is mentioned in failing rather than passing test cases (Section 3.4); combine the three scores (edep, cdep, dyn) into the score fixme, which determines a global *ranking* of expressions (Section 3.5); enumerate possible *fixing actions* for the expressions with highest fixme score (Section 3.6); generate *candidate fixes* by injecting the fix actions into the faulty routine (Section 3.7); the candidate fixes that pass all the regression test suite are considered *valid* (Section 3.8).

line 14: *forth* moves the cursor to the first position, which is valid for *put_left*.

## 2.3 Model-Based Fixing at Work

How do model-based techniques, implemented in AutoFix-E, perform on the two errors shown? The error in the invocation of *put_left* has a characterization in terms of public queries and state invariants, hence AutoFix-E also produces a correct fix, equivalent to the one from AutoFix-E2.

Model-based techniques, however, can correct the other error, in the invocation of *go_i_th*, only for the specific instance exposed by the test case where $index = count + 1$, that is when *after* holds. Based on this, a possible partial fix consists in adding **if** *after* **then** *back* **end** as first instruction on line 5. This fix is not only partial but also unlikely to be generated in practice, because it modifies code which is several instructions away from where the contract violation occurs, but AutoFix-E's heuristics favor fixes that are local to restrict the search space. As shown above, code-based techniques do not suffer these limitations.

## 3 Code-Based Fixing

This section describes how code-based fixing works; Figure 2 depicts the main steps of the process. All the running examples refer to Listing 1.

Code-based fixing works on Eiffel classes equipped with contracts [16]: preconditions, postconditions, and class invariants. Each contract element consists of one or more *clauses*; for example, *move_item*'s precondition on line 13 has two clauses: $v \neq \textbf{Void}$ and $has(v)$. The contracts of a class constitute its executable specification, hence provide a way to determine functional *errors* in the implementation: a routine called in a state not satisfying its precondition, terminating in a state not satisfying its postcondition or violating the class invariant, or reaching an intermediate assertion not satisfied.

## 3.1 Test-Case Generation

Every session of code-based fixing starts by collecting information about the runtime behavior of the routine under fix. The raw form of such information is a collection of *test cases*, each a sequence of object creations and routine invocations on the objects. A test case is *passing* if it does not violate any contract and *failing* otherwise. Two failing test cases correspond to the same error if they violate the same contract clause at the same program location; this assumption is reasonable, given that different clauses of the same contract are usually orthogonal.

Code-based fixing takes a set $T$ of test cases as input, and uses them for dynamic analysis (Section 3.4) and fix validation (Section 3.8). $P$ and $F$ respectively denote the sets of all passing and failing test cases in $T$. $F^{j,c}$ denotes a set of failing test cases violating the clause $c$ at program location $j$. For example, the set of test cases violating *put_left*'s precondition in *move_item* is denoted by $F^{14,(\mathbf{not}\ before)}$. Each session of code-based fixing targets a single fault.

The rest of the code-based fixing process is independent of whether the test cases $T$ are generated automatically or written manually. AutoFix-E2 uses the random testing framework AutoTest [17] developed in previous work of ours. The use of AutoTest makes the fixing process in AutoFix-E2 completely automatic. The experiments described in Section 4 demonstrate that the test cases generated by AutoTest are suitable inputs to AutoFix-E2 and support the generation of effective fixes without any human intervention.

## 3.2 Predicates, Expressions, and States

Evidence takes the form of boolean *predicates*, built by combining *expressions* extracted from the program text and the violated contract clause. The evaluation of a predicate at a program location gives a *component* of the program state at that location. Sections 3.3 and 3.4 rank components according to their "suspiciousness" of being responsible for the occurrence of an error.

### 3.2.1 Expressions

For a routine $r$ and a violated assertion clause $c$, $E_{r,c}$ denotes the set of non-constant *expressions* (of any type) which appear in $r$'s body or in $c$. For example, $E_{before, index\ >1}$ for routine *before* is {**Result**, *index*, *index* = 0, *index* > 1}. $\mathbb{E}_{r,c}$ extends the set $E_{r,c}$ of expressions by *unfolding* [18]: $\mathbb{E}_{r,c}$ includes all elements in $E_{r,c}$ and, for every $e \in E_{r,c}$ of reference type $t$ and for every argument-less query $q$ applicable to objects of type $t$, $\mathbb{E}_{r,c}$ also includes the expression $e.q$. Continuing the example, $\mathbb{E}_{before, index\ >1} = E_{before, index\ >1}$ because all the expressions in $E_{before, index\ >1}$ are of primitive type (integer or boolean).

### 3.2.2 Predicates

The set $\mathbb{P}_{r,c}$ of boolean *predicates* generated for $r$ and $c$ contains the following elements:

**Boolean expressions**: $b$, for boolean $b \in \mathbb{E}_{r,c}$ of boolean type;

**Voidness checks**: $e = \mathbf{Void}$, for every $e \in \mathbb{E}_{r,c}$ of reference type;

**Integer comparisons**: $e \sim e'$, for every $e \in \mathbb{E}_{r,c}$ of integer type, every $e' \in \mathbb{E}_{r,c} \setminus \{e\} \cup \{0\}$ also of integer type, and every $\sim$ in $\{=, <, \leq\}$;

**Complements**: $\neg p$, for every $p \in \mathbb{P}_{r,c}$.

For example, $\mathbb{P}_{before, index\, >1}$ contains **Result**, **not Result**, $index \sim 0$ and $index \sim 1$ ($\sim \in \{=, \neq, <, <=, >, >=\}$).

### 3.2.3 State Components

A test case $t \in T$ describes a sequence $\mathsf{loc}(t) = \ell_1, \ell_2, \ldots$ of executed program locations. For an expression $e$ and a location $\ell \in \mathsf{loc}(t)$, $[\![e]\!]_t^\ell$ is the value of $e$ at $\ell$ in $t$, if $e$ can be evaluated at $\ell$.

The evaluation of predicate $p$ at location $\ell$ defines the triple $\langle \ell, p, v \rangle$, where $v$ is the value $[\![p]\!]_t^\ell$ for some test case $t$ which reaches $\ell$; a test case $t$ may define multiple triples with the same $\ell$, if $\ell$ appears more than once in $\mathsf{loc}(t)$. $\mathsf{comp}(T)$ denotes all the triples $\langle \ell, p, v \rangle$ defined by the tests in the set $T$; they are the *components* of the program state during the tests. In the running example, every test case reaching location 6 defines $\langle 6, v = \mathbf{Void}, \mathbf{False} \rangle$—because the precondition guarantees $v \neq \mathbf{Void}$—but does not define any triple $\langle 6, \mathbf{Result}, v \rangle$—because **Result** is not a variable in the scope of *move_item*.

Sections 3.3–3.5 show how to *rank* components according to heuristics which take into account static and dynamic measures. The ranking heuristic fixme summarizes various sources of evidence; a triple $\langle \ell, p, v \rangle$ appearing high in the ranking indicates that an error is likely to have its origin at location $\ell$ when predicate $p$ evaluates to $v$. Correspondingly, the fixes generated automatically try to change the value of $p$ at $\ell$ whenever it is $v$ (Section 3.6).

## 3.3 Static Analysis

Static analysis extracts evidence from the program text independently of the runtime behavior: control dependence measures the distance, in terms of number of instructions, between two program locations; expression dependence measures the syntactic similarity between two predicates. We use control dependence to estimate the proximity of a location to where a failure is triggered; then, we further differentiate among expressions evaluated at nearby program locations according to a simple syntactic measure of similarity between each expression and the violated contract clause. Such a lightweight static analysis is sufficient for code-based fixing to work, given that the primary source of evidence comes from dynamic analysis (Section 3.4) anyway.

### 3.3.1 Control Dependence

For two program locations $\ell_1, \ell_2$, write $\ell_1 \rightsquigarrow \ell_2$ if $\ell_1$ and $\ell_2$ belong to the same routine and there exists a directed path from $\ell_1$ to $\ell_2$ on the control-flow graph of the routine's body; otherwise, $\ell_1 \not\rightsquigarrow \ell_2$. The *control distance* $\mathsf{cdist}(\ell_1, \ell_2)$ of two program locations is the length of the shortest directed path from $\ell_1$ to $\ell_2$ on the control-flow graph if $\ell_1 \rightsquigarrow \ell_2$, and $\infty$ if $\ell_1 \not\rightsquigarrow \ell_2$. For example, $\mathsf{cdist}(8, 12) = 4$ in Listing 1.

Correspondingly, the *control dependence* $\mathsf{cdep}(\ell, \jmath)$ is the normalized score:

$$\mathsf{cdep}(\ell, \jmath) \;=\; 1 - \frac{\mathsf{cdist}(\ell, \jmath)}{\max\{\mathsf{cdist}(\lambda, \jmath) \mid \lambda \in r \text{ and } \lambda \rightsquigarrow \jmath\}}$$

for $\ell \rightsquigarrow \jmath$, and 0 for $\ell \not\rightsquigarrow \jmath$. We use control dependence to rank locations according to proximity to the location of failure.

### 3.3.2 Expression Dependence

For an expression $e$, define the set $\mathsf{sub}(e)$ of its sub-expressions as follows:

- $e \in \mathsf{sub}(e)$;
- if $e' \in \mathsf{sub}(e)$ is a query call of the form $t.q\,(a_1, \ldots, a_m)$ for $m \geq 0$, then $t \in \mathsf{sub}(e)$ and $a_i \in \mathsf{sub}(e)$ for all $1 \leq i \leq m$.

This definition also accommodates infix operators (such as boolean connectives and arithmetic operators), which are just syntactic sugar for query calls; for example $a$ and $b$ are both sub-expressions of $a + b$, a shorthand for $a.\mathit{plus}\,(b)$. Also, unqualified query calls are treated as qualified call on the implicit target **Current**.

The *expression proximity* $\mathsf{eprox}(e_1, e_2)$ of two expressions $e_1, e_2$ measures how similar $e_1$ and $e_2$ are in terms of shared sub-expressions: $\mathsf{eprox}(e_1, e_2) = |\mathsf{sub}(e_1) \cap \mathsf{sub}(e_2)|$ For example, $\mathsf{eprox}(i \leq \mathit{count}, 0 \leq i \leq \mathit{count} + 1)$ is 2, corresponding to the shared sub-expressions $i$ and $\mathit{count}$. The larger the expression proximity between two expressions is, the more similar they are.

Correspondingly, the *expression dependence* $\mathsf{edep}(p, c)$ is the normalized score measuring the amount of evidence that $p$ and $c$ are syntactically similar:

$$\mathsf{edep}(p, c) \;=\; \frac{\mathsf{eprox}(p, c)}{\max\{\mathsf{eprox}(\pi, c) \mid \pi \in \mathbb{P}_{r,c}\}}$$

In routine *before*, for example, $\mathsf{edep}(\mathit{index}, \mathit{index} = 0)$ is $1/3$ because $\mathsf{edep}(\mathit{index}, \mathit{index} = 0) = 1$ and $\mathit{index} = 0$ itself has the maximum expression proximity to $\mathit{index} = 0$. We use expression dependence to rank expressions according to similarity to the contract violated by a failure. Expression dependence is meaningful only for expressions evaluated in the same local environment (that is, with strong control dependence), where the same syntax is likely to refer to identical program elements.

## 3.4 Dynamic Analysis

Dynamic analysis extracts evidence from test cases in the form of score associated to every state component generated. The higher the score $\mathsf{dyn}\langle \ell, p, v \rangle$ a component $\langle \ell, p, v \rangle$ receives, the stronger the runtime behavior suggests that an error originates at location $\ell$ when predicate $p$ evaluates to $v$.

### 3.4.1 Principles to Compute the Score

Consider an error violating the contract clause $c$ at location $\jmath$ in some routine $r$. Let $P_r$ be a set of passing test cases exercising routine $r$, $F_r^{\jmath,c}$ a set of failing test cases exposing the same error, and $T_r^{\jmath,c}$ the union $P_r \cup F_r^{\jmath,c}$. $\mathsf{comp}(T_r^{\jmath,c})$ is then the set of components describing correct and incorrect behavior of $r$.

For a test case $t \in T_r^{\jmath,c}$ and a component $\langle \ell, p, v \rangle$ such that $\ell$ is a location in $r$'s body, write $\langle \ell, p, v \rangle \in t$ if $t$ reaches location $\ell$ at least once and $p$ evaluates to $v$ there:

$$\langle \ell, p, v \rangle \in t \quad \text{iff} \quad \exists \ell_i \in \mathsf{loc}(t), \ell = \ell_i, \text{ and } v = [\![p]\!]_t^{\ell_i}$$

For every test case $t \in T_r^{\jmath,c}$ such that $\langle \ell, p, v \rangle \in t$, $\sigma(t)$ denotes its contribution to the score of $\langle \ell, p, v \rangle$: a large $\sigma(t)$ denotes evidence that $\langle \ell, p, v \rangle$ is a likely "source" of error if $t$ is a failing test case, and evidence against it if $t$ is passing.

Section 3.4.2 builds a function $\sigma$ according to the following principles:

(a) If there is at least one failing test case $t$ such that $\langle \ell, p, v \rangle \in t$, the overall score assigned to $\langle \ell, p, v \rangle$ must be positive: the evidence provided by failing test cases cannot be canceled out completely.

(b) The magnitude of each failing (resp. passing) test case's contribution $\sigma(t)$ to the score assigned to $\langle \ell, p, v \rangle$ decreases as more failing (resp. passing) test cases for that component are available: the evidence provided by the first few test cases is crucial, while repeated outcomes carry a lower weight.

(c) The evidence provided by one failing test case is stronger than the evidence provided by one passing test case.

The first two principles are after Wong et al.'s "Heuristic III" [23], which experiments by the same authors showed to yield better fault localization accuracy than most alternative approaches. According to these principles, components appearing only in failing test cases are more likely to be fault causes.

Our dynamic analysis assigns scores according to the same basic principles as Wong et al.'s, but with differences suggested by the ultimate goal of automatic fixing: our score ranks state components rather than program locations, and assigns weight to test cases differently. Contracts significantly help find the location responsible for a fault: in many cases, it is proximate to where the contract violation occurred; on the other hand, automatic fixing requires to gather information not only about the location but also about the state "responsible" for the fault. This observation prompted us to apply the fault localization principles to state components.

### 3.4.2 Score from Dynamic Analysis

Assume an arbitrary order on the test cases and let $\sigma(t)$ be $\alpha^i$ for the $i$-th failing test case $t$ and $\beta \alpha^i$ for the $i$-th passing test case. Selecting $0 < \alpha < 1$ decreases the contribution of each test case exponentially, which meets principle (b); then, selecting $0 < \beta < 1$ fulfills principle (c).

The evidence provided by each test case adds up:

$$\mathsf{dyn}\langle \ell, p, v \rangle = \gamma + \sum \{\sigma(u) \mid u \in F_r^{\jmath,c}\} - \sum \{\sigma(v) \mid v \in P_r\}$$

for some $\gamma \geq 0$; the chosen ordering is immaterial. We compute the score with the closed form of geometric progressions:

$$\#\mathsf{p}\langle \ell, p, v \rangle = |\{t \in P_r \mid \langle \ell, p, v \rangle \in t\}|$$
$$\#\mathsf{f}\langle \ell, p, v \rangle = |\{t \in F_r^{j,c} \mid \langle \ell, p, v \rangle \in t\}|$$
$$\mathsf{dyn}\langle \ell, p, v \rangle = \gamma + \frac{\alpha}{1-\alpha}\left(1 - \beta + \beta\alpha^{\#\mathsf{p}\langle \ell,p,v \rangle} - \alpha^{\#\mathsf{f}\langle \ell,p,v \rangle}\right)$$

where $\#\mathsf{p}\langle \ell, p, v \rangle$ and $\#\mathsf{f}\langle \ell, p, v \rangle$ are the number of passing and failing test cases that determine the component $\langle \ell, p, v \rangle$. It is straightforward to prove that $\mathsf{dyn}\langle \ell, p, v \rangle$ is positive if $\#\mathsf{f}\langle \ell, p, v \rangle \geq 1$, for every $0 < \alpha, \beta < 1$, hence the score meets principle (a) as well. Some empirical evaluation suggested to set $\alpha = 1/3$, $\beta = 2/3$, and $\gamma = 1$ in the current implementation of AutoFix-E2.

## 3.5 Combining Static and Dynamic Analysis

The final output of the analysis phase combines static and dynamic analysis to assign a "suspiciousness" score $\mathsf{fixme}\langle \ell, p, v \rangle$ to every state component $\langle \ell, p, v \rangle$.

Expression dependence and control dependence are both *ratios*, and the score from dynamic analysis is essentially a sum of fractional values. This suggests [3] to combine the three scores by harmonic mean:

$$\mathsf{fixme}\langle \ell, p, v \rangle = \frac{3}{\mathsf{edep}(p,c)^{-1} + \mathsf{cdep}(\ell,j)^{-1} + \mathsf{dyn}\langle \ell, p, v \rangle^{-1}}$$

The current choice of parameters $\alpha, \beta, \gamma$ makes the dynamic score $\mathsf{dyn}\langle \ell, p, v \rangle$ dominant in determining the overall score $\mathsf{fixme}\langle \ell, p, v \rangle$: while expression and control dependence vary between 0 and 1, the dynamic score has minimum 2/3 (for zero failing test cases and indefinitely many passing) and maximum 3/2 (for zero passing test cases and indefinitely many failing). This range difference is consistent with the principle that dynamic analysis gives the primary source of evidence, whereas the less precise evidence provided by static analysis is useful to discriminate among components with similar dynamic behavior.

## 3.6 Fixing Actions

Consider a component $\langle \ell, p, v \rangle$ with a high evidence score $\mathsf{fixme}\langle \ell, p, v \rangle$. $\langle \ell, p, v \rangle$ induces a number of possible *actions* (instructions) which try to avoid using the value $v$ of $p$ at $\ell$. The actions may either modify $p$ directly (Section 3.6.2) or change the usage of $p$ in the instruction at $\ell$ (Section 3.6.3).

### 3.6.1 Derived Expressions

Expressions of boolean and integer type are modified according to standard patterns which may reverse common sources of mistakes—such as "off-by-one" errors. For an expression $e$, the set $\mathsf{ederiv}(e)$ includes:

- if $e$ is of boolean type, the constants **True** and **False**, and the expression **not** $e$;

- if $e$ is of integer type, the constants $0, 1, -1$, and the expressions $e + 1$ and $e - 1$.

### 3.6.2 Expression Modification

One way to change a state component is to directly modify the expression of that component. An expression $e$ is *modifiable* at $\ell$ if: $e$ is of reference type; or $e$ is of integer type and the assignment $e := 0$ can be executed at $\ell$; or $e$ is of boolean type and the assignment $e := \textbf{True}$ can be executed at $\ell$. For example, *index* is modifiable everywhere in routine *move_item* because it is an attribute of the enclosing class; in routine *go_i_th*, instead, $i$ is not modifiable anywhere because arguments are read-only in Eiffel.

Since an expression in a state component may not be directly modifiable, we also consider sub-expressions. The definition of sub-expression (Section 3.3.2) induces a partial order $\preceq$: $e_1 \preceq e_2$ iff $e_1 \in \mathsf{sub}(e_2)$; correspondingly, it defines the *largest* expressions in a set. For example, the largest expressions of integer type in $\mathsf{sub}(idx < index \textbf{ or } after)$ are *idx* and *index*. A pair $\langle \ell, p \rangle$ determines the set $\mathsf{targ}\langle \ell, p \rangle$ of *target* expressions: $\mathsf{targ}\langle \ell, p \rangle$ includes the largest expressions among $p$'s sub-expressions $\mathsf{sub}(p)$ that are modifiable at $\ell$. For example, $\mathsf{targ}\langle 13, idx > \textbf{Current}.index \rangle$ on Listing 1 includes the integer expressions **Current**.*index* and *idx*, but no reference (**Current** is a sub-expression of **Current**.*index*) or boolean expression (*idx* $>$ **Current**.*index* is not modifiable according to the definition).

Finally, populate the set $\mathsf{emod}\langle \ell, p \rangle$ of expression modifications induced by the component $\langle \ell, p, v \rangle$ as follows:

- for $e \in \mathsf{targ}\langle \ell, p \rangle$ of boolean or integer type and every derived expression $d \in \mathsf{ederiv}(e)$, include $e := d$ in $\mathsf{emod}\langle \ell, p \rangle$;

- for $e \in \mathsf{targ}\langle \ell, p \rangle$ of reference type, if $e.c\,(a_1, \ldots, a_n)$ is a call to a command (procedure) $c$ executable at $\ell$, include $e.c\,(a_1, \ldots, a_n)$ in $\mathsf{emod}\langle \ell, p \rangle$.

In the running example, $\mathsf{emod}\langle 13, idx > \textbf{Current}.index \rangle$ includes assignments of 0, 1 and $-1$ to *idx* and *index*, and unit increments and decrements of the same variables.

### 3.6.3 Expression Replacement

There are cases where expression modification is infeasible or undesirable. For example, expression $i$ in routine *go_i_th* does not have any modifiable sub-expression. In such situations, expression replacement directly substitutes the usage of expressions in instructions.

Every location $\ell$ labels either a primitive instruction (an assignment or a routine call) or a boolean condition (the branching condition of an **if** instruction or the exit condition of a **loop**). Correspondingly, define the set $\mathsf{sub}(\ell)$ of sub-expressions of a *location* $\ell$ as follows:

- if $\ell$ labels a boolean condition $b$ then $\mathsf{sub}(\ell) = \mathsf{sub}(b)$;

- if $\ell$ labels an assignment $v := e$ then $\mathsf{sub}(\ell) = \mathsf{sub}(e)$;

- if $\ell$ labels a routine call $t.c\,(a_1, \ldots, a_n)$ then $\mathsf{sub}(\ell) = \bigcup \{\mathsf{sub}(a_i) \mid 1 \leq i \leq n\}$.

A pair $\langle \ell, p \rangle$ determines the set $\mathsf{erepl}\langle \ell, p \rangle$ of instructions with replaced expressions as follows: for each expression $e$ among the largest sub-expressions of boolean or integer type in $\mathsf{sub}(p)$, if $e \in \mathsf{sub}(\ell)$ then include $\ell[e \mapsto e']$ in $\mathsf{erepl}\langle \ell, p \rangle$, for every $e' \in \mathsf{ederiv}(e)$. $\ell[e \mapsto e']$ denotes the instruction obtained by replacing every occurrence of $e$ at location $\ell$ with $e'$; if $\ell$ labels a boolean condition, $\ell[e \mapsto e']$ denotes the whole *instruction* (conditional or loop) but $e'$ replaces $e$ only in the boolean condition.

In the continued example, $\mathsf{erepl}\langle 13, idx > index\rangle$ includes $go\_i\_th\ (idx - 1)$, $go\_i\_th\ (idx + 1)$, $go\_i\_th\ (0)$, $go\_i\_th\ (1)$, and $go\_i\_th\ (-1)$; $\mathsf{erepl}\langle 13, idx + 1 > index\rangle$, however, is empty because the two largest integer sub-expressions in $idx + 1 > index$ are $idx + 1$ and $index$, none of which is a sub-expression of $idx$ in $go\_i\_th\ (idx)$. In the same routine, $\mathsf{erepl}\langle 9, found\rangle$ includes the conditional instructions **if not**(**not** *found*) **then** *forth* **end**, **if True then** *forth* **end**, and **if False then** *forth* **end**.

### 3.7 Fix Candidate Generation

At this point, for any "suspicious" state component $\langle \ell, p, v \rangle$ we can generate actions that change the value (3.6.2) or the usage (3.6.3) of $p$ at $\ell$. Each such action generates a *candidate fix* if injected at location $\ell$. The injection consists of first selecting a *fix schema* (3.7.1), then *instantiating* the schema with $p$ and an action derived from $p$ (3.7.2).

#### 3.7.1 Fix Schemas

We use the same fix schemas used for model-based fixing [21] shown in Table 1.

| (a) | (b) | (c) | (d) |
|---|---|---|---|
| *snippet* | **if** *fail* **then** | **if not** *fail* **then** | **if** *fail* **then** |
| *old_stmt* | *snippet* | *old_stmt* | *snippet* |
| | **end** | **end** | **else** |
| | *old_stmt* | | *old_stmt* |
| | | | **end** |

Table 1: Fix schemas.

#### 3.7.2 Schema Instantiation

For a state component $\langle \ell, p, v \rangle$ determined by the passing test cases $P_r$ of routine $r$ and the failing test cases $F_r^{\jmath,c}$ violating the contract clause $c$ at location $\jmath$, instantiate each of the schemas in Table 1 as follows:

*fail* takes $p = v$, the component's predicate and value.

*snippet* takes any value in $\mathsf{emod}\langle \ell, p \rangle \cup \mathsf{erepl}\langle \ell, p \rangle$ (defined in Sections 3.6.2–3.6.3).

*old_stmt* is the instruction at location $\ell$.

The instantiated schema replaces the instruction at position $\ell$ in routine $r$; the modified routine is a *candidate fix*.

In the running example, the component $\langle 13, idx > index, \textbf{True}\rangle$ leads to several candidate fixes. Schema (b) with the component's predicate $idx > index$ as *fail*, the expression modification $idx := idx - 1$ as *snippet*, and the original instruction $go\_i\_th\ (idx)$ as *old_stmt* produces a correct fix. A different combination, which also produces a correct fix, is schema (d) with *fail* using the component's predicate $idx > index$, the instruction with the expression replacement $go\_i\_th\ (idx - 1)$ as *snippet*, and the original instruction $go\_i\_th\ (idx)$ as *old_stmt*.

### 3.8 Validation of Candidates

The generation of candidate fixes involves the application of several heuristics and is essentially "best effort": there is no *a priori* guarantee that the candidates actually fix the program. Each candidate fix must pass a validation phase which determines whether its deployment removes the erroneous behavior under consideration. The validation phase runs each candidate fix through the full set of passing and failing test cases. A fix is *validated* if it passes all the previously failing test cases $F_r^{?,c}$ and it still passes the original passing test cases $P_r$. In general, more than one candidate fix may pass the validation phase; AutoFix-E2 ranks all valid fixes according to the score of the state component that originated the fix and submits the top 15 to the user, who is ultimately responsible to decide whether to deploy any of them.

The correctness of a program is defined relative to its specification; in the case of automated program fixing, the validated fixes are only as good as the contracts. For example, routine *move_item* lacks a postcondition, therefore the simple candidate fix which unconditionally adds the assignment $idx := 1$ before the call to $go\_i\_th$ is validated despite being obviously inappropriate. In spite of these limitations in principle, the experiments in Section 4 show that the available contracts are often good enough in practice, so that AutoFix-E2 suggests *proper* fixes—correct not only according to the contracts available but also to the intuitive expectations of developers—in the large majority of cases where it can validate some fixes. Improving the quality of the contracts is a related effort which can also greatly benefit from automation [20] and whose results boost the effectiveness of automated program fixing.

## 4 Experimental evaluation

### 4.1 Experimental Setup

All the experiments ran on a Windows 7 machine with a 2.66 GHz Intel dual-core CPU and 4 GB of memory. On average, AutoFix-E2 ran for 7.6 minutes for each fault.

#### 4.1.1 Selection of Faults

The experiments include faults from two sources: data structure classes from commercial libraries, and an implementation of a library to manipulate text documents developed as student project.

**Data structure libraries.** Table 2 lists the 15 classes from the Eiffel-Base [7] (rev. 507) and Gobo [9] (rev. 79072) libraries used in the experiments;

the table reports the length in lines of code (LOC), the total number of routines (#R) and boolean queries (#B) of each class, and the number of faults (#F) considered in the experiments. This selection of faults combines 13 faults used in the evaluation of model-based fixing [21] with 51 new faults recently found by AutoTest. We did not re-use the remaining 29 faults used in [21] because they are not reproducible in the latest revision of the libraries.

Table 2: EiffelBase and Gobo classes.

| CLASS | LOC | #R | #B | #F |
|---|---|---|---|---|
| *ACTIVE_LIST* | 2162 | 139 | 19 | 2 |
| *ARRAY* | 1464 | 101 | 11 | 9 |
| *ARRAYED_CIRCULAR* | 1910 | 133 | 25 | 3 |
| *ARRAYED_SET* | 2345 | 146 | 18 | 5 |
| *DS_ARRAYED_LIST* | 2762 | 166 | 9 | 5 |
| *DS_HASH_SET* | 3076 | 169 | 10 | 1 |
| *DS_LINKED_LIST* | 3434 | 160 | 8 | 5 |
| *HASH_TABLE* | 2036 | 118 | 19 | 2 |
| *INTEGER_32* | 1115 | 99 | 5 | 1 |
| *LINKED_LIST* | 2000 | 109 | 16 | 1 |
| *LINKED_PRIORITY_QUEUE* | 2374 | 125 | 17 | 1 |
| *LINKED_SET* | 2352 | 122 | 16 | 5 |
| *REAL_64* | 839 | 72 | 4 | 1 |
| *SUBSET_STRATEGY_HASHABLE* | 543 | 33 | 0 | 4 |
| *TWO_WAY_SORTED_SET* | 2868 | 141 | 18 | 19 |
| **Total** | 31280 | 1833 | 195 | 64 |

**A library to manipulate text documents.** The second part of the evaluation targets a library to manipulate text documents and convert them into HTML and LaTeX. The library models entities such as formatted text, lists, tables, and images; it has been implemented as a student project of the Software Architecture course held in the spring semester 2010 at ETH Zurich. Table 3 lists the 3 classes of the library used in the experiments, with the same statistics as in Table 2. Compared to EiffelBase and Gobo, the text document library's classes have a more primitive interface, with very few boolean queries (31 of the 32 boolean queries of class *FILE_NAME* are inherited from the library class *STRING*, hence they are mostly unrelated to the specific semantics of *FILE_NAME*) and less detailed contracts; therefore, they are representative of less mature software with functionalities complex to specify formally. AutoTest detected 9 faults (#F) in the classes: 5 precondition violations, 3 intermediate assertion violations, and 1 call on void target (null pointer dereference).

Table 3: Document manipulation library classes.

| CLASS | LOC | #R | #B | #F |
|---|---|---|---|---|
| *FILE_NAME* | 4297 | 258 | 32 | 2 |
| *HTML_TRANSLATOR* | 1148 | 83 | 0 | 1 |
| *LATEX_TRANSLATOR* | 1269 | 90 | 0 | 6 |
| **Total** | 6714 | 431 | 32 | 9 |

### 4.1.2 Selection of Test Cases

All the experiments used test cases generated automatically by AutoTest; this demonstrates complete automation of the whole debugging process and minimizes the potential bias introduced by experimenters. AutoTest produced an average of 25 passing and 11 failing test cases for each fault.

## 4.2 Experimental Results

### 4.2.1 Data Structure Libraries

Table 4 summarizes the results of the experiments on the data structure libraries: the number #F of faults in each category, the faults fixed with model-based techniques using AutoFix-E, and those fixed with code-based techniques using AutoFix-E2. The count of valid fixes only includes those which are *proper*, that is which manual inspection confirmed to be adequate beyond the correctness criterion provided by the contracts and tests available. The faults fixed by AutoFix-E2 are a superset of those fixed by AutoFix-E; when both tools succeeded, they produced equivalent fixes (with possibly negligible syntactic differences). We refrained from injecting more bugs in EiffelBase and Gobo—as it is customary in evaluating fault localization techniques—in order to have an evaluation that only deals with real bugs found in production software.

Of the 50 faults not fixed, about 25 expose design errors, rather than mere programming errors: for example, several of the faults point to inconsistencies in the inheritance hierarchy of the library. Another 19 faults originate from incorrect or incomplete contracts, such as weak class invariants that let objects reach inconsistent states. The remaining 6 faults are of various type, including some non-functional properties. To our knowledge, automatically fixing most of these "deep" errors is beyond the capabilities of any existing automatic program fixing technique.

Table 4: Faults fixed in EiffelBase and Gobo classes.

| Type of fault | # F | Model | Code |
|---|---|---|---|
| Precondition violation | 22 | 10 (45%) | 12 (54%) |
| Postcondition violation | 30 | 0 (0%) | 2 (6%) |
| Call on void target | 7 | 0 (0%) | 0 (0%) |
| Intermediate assertion violation | 5 | 0 (0%) | 0 (0%) |
| **Total** | **64** | **10(15%)** | **14(22%)** |

The results show that code-based techniques constitute a significant improvement over model-based techniques. Even if model-based techniques perform already quite satisfactorily on data structure implementations, due to the high quality of the queries available in their interfaces, code-based fixing succeeded with 4 more errors (40% improvement). Most of the errors where code-based fixing succeeds and model-based techniques fail are indicative of subtle bugs with non-obvious fixes. Three are precondition violations: one is described in Section 2; the other two—from class *DS_HASH_SET* and *HASH_TABLE*—are similar in that the fix requires to reference a local variable rather than public queries. The other fault is a postcondition violation, which model-based techniques cannot handle as it requires a fix in a location different than where the

Listing 2: Routine *visit_table* fixed by AutoFix-E2.

```
1   visit_table (a_table: TABLE)
2     local s: STRING; i: INTEGER
3     do
4        packages.extend ("tabulary")
5        create s.make_empty
6        if a_table.count >0 then -- Added by AutoFix-E2
7          from i := 1
8          until i > a_table.column_count loop
9             -- Append to 's' the table's content,
10            -- in LaTeX form
11         end
12       end -- Added by AutoFix-E2
13       ...
14       s.prepend ("|") ; s.append ("|")
15       open_environment ("tabulary", s)
16       ...
17    end
```

violation occurs (i.e., at the end of the routine's body).

### 4.2.2 Text Document Manipulation Library

The second set of experiments tried to determine if code-based techniques can successfully tackle software beyond well-engineered data structure implementations. AutoFix-E2 built valid fixes for 5 of the 9 faults in the document library: one in each of the classes *FILE_NAME* and *HTML_TRANSLATOR*, and 3 in the class *LATEX_TRANSLATOR*. In comparison, AutoFix-E only fixed one of the faults, which AutoFix-E2 also fixed; manual inspection confirmed the expectation that model-based fixing fails whenever the fault conditions cannot be characterized using only boolean queries—the case for nearly all the errors in the text document library.

As an example from these experiments, Listing 2 shows the essential parts of a routine *visit_table* fixed by AutoFix-E2. *visit_table* converts data in table form, passed as argument *a_table*, into LaTeX. To this end, it first opens a "tabulary" environment (line 4); then, the loop on lines 7–11 converts the content of the various columns into the string *s*; finally, it adds delimiters to the table (line 14), and stores the content of *s* in the "tabulary" environment (line 15). The loop fails when the table is empty, because the query *column_count* of *a_table* has a precondition *count*>0. The fix wraps a conditional statement (lines 6–12) around the loop; correspondingly, an empty table becomes an empty LaTeX table as appropriate. This example gives an idea of the kinds of fixes generated in the second set of experiments, and how code-based techniques can be successful on them.

### 4.2.3 Overall Performance of Code-Based Fixing

> *In the experiments, code-based techniques fixed 19 errors, 73% more than model-based techniques.*

## 4.3 Threats to Validity

Some threats may limit the generalizability of the results:

- The choice of using only automatically generated test cases may affect the performance and quality of the results. On the other hand, if code-based fixing works well also with manually written test cases it can be applicable to more software.

- The evaluation uses real software and real errors made by programmers, but it could target even more classes of diverse application domains. A larger-scale thorough evaluation belongs to future work.

- Our notion of correctness is relative to the available contracts. Correspondingly, the quality of the contracts may affect the quality of fixes produced, but we do not know to what extent this holds for the classes used in the experiments.

## 5 Related Work

This section summarizes the most relevant related work in two areas: fault localization and automated program fixing.

### 5.1 Fault Localization

Fault localization is the process of locating erroneous statements in a program. Several suggested solutions to this problem use heuristics based on code coverage (e.g., [13, 19]) or program states (e.g., [11, 24]).

**Code coverage.** Code coverage metrics have been used to rank instructions based on their likelihood to trigger failures. [13], for example, introduces the notion of *failure rate*: based on a large number of test cases, an instruction has a high failure rate if it is executed more often in failing test cases than in passing test cases. A block of code is then "suspicious" of being faulty if it includes many instructions with high failure rate; [13] suggests to visualize the failure rates with colors and brightness, and implements the scheme in the tool Tarantula. [19] proposes a fault localization technique named *nearest neighbor*. The nearest neighbor of a given faulty test case is the passing test case in a test suite which is most similar to the failing test case. Removing all the instructions mentioned in the nearest neighbor from the faulty test produces a smaller set of instructions; these are the candidates to be responsible for the fault under consideration. Several other authors have extended code coverage techniques for fault localization. For example, [25] addresses the propagation of infected program states; [15] relies on a model-based approach; and [23] performs an extensive comparison of variants of fault localization techniques and outlines general principles behind them. [4] discusses the limitations of using only state invariants for fault localization, a limitation present in model-based fixing techniques but removed with the code-based approach.

**Program states.** The application of code coverage techniques produces a set of instructions likely to be responsible for failure; programmers still have to examine each instruction to understand where the problem is. Fault localization techniques based on program states aim at alleviating this task. [11], for example, requires programmers to insert check points in the program to mark "points of interest". Then, a dynamic analysis similar to [13]—but applied to program states rather than locations—identifies a set of suspicious states. Such a state-based analysis is finer-grained than those based only on code coverage; furthermore, the usage of check points introduces more flexibility to skip uninteresting parts of the computation, for example repeated iterations of a loop. *delta debugging* [24] addresses similar issues: isolating the variables, and their values, relevant to a failure by analyzing the state difference between passing and failing test cases. Most fault localization techniques target each fault individually, hence they perform poorly when multiple bugs interact and must be considered together. To address such scenarios, [14] introduces a technique that separates the effects of multiple faults and identifies predictors associated with each fault.

**Fault localization in code-based fixing.** The code-based program fixing techniques of the present paper also exploit fault localization techniques. To generate fixes completely automatically, however, fault localization must be sufficiently precise to suggest only a limited number of "suspicious" instructions. In our case, the usage of contracts help to restrict the search to the boundaries of the routine where a contract violation occurs. Then, the combination of static and dynamic analysis techniques that rank state components within routines produces fault localization sufficiently accurate for fixing faults automatically.

## 5.2 Automated Program Fixing

This section reviews the most significant contributions to automated fixing of source code. The related work section in our previous work [21] also describes different approaches working at runtime on the compiled binary.

[12] presents BugFix, a tool that helps developers fix bugs by suggesting patches. Their approach uses machine-learning techniques, which can work without annotations such as contracts. BugFix learns existing fixes in the form of association rules, and it tries to apply the rules learned to new bugs. Users can provide feedback—in the form of new examples of correct fixes or validations of suggestions provided by the tool—which ameliorates the quality of the suggestions provided over time.

Other authors apply genetic algorithms to generate fixes automatically. [1] uses a co-evolutionary scheme where an initially faulty program and some test cases compete to evolve the program into one that satisfies its formal specification. [22] describes a technique, based on genetic algorithm, that takes a program, a set of successful test cases, and one failing test case. After rounds of evolution, the program changes into one that passes all test cases (including the failing test case). While [22]'s results are significant, as they can patch real programs of non-trivial size, the role played by evolutionary techniques is not entirely clear: as pointed out also in [2], the experiments of [22] span only a limited number of generations (about 10), which suggests that the genetic algorithm performs only a very limited search in the space of possible solutions. Another limitation of [22] resides in its sensitivity to the quality (and size) of the provided test suite, an effect which is much less relevant in our approach where random testing techniques can generate a suitable test suite automatically.

[10] presents a technique that compares two program states at a faulty location in the program. Unlike all other approaches to program fixing to date,

[10] computes program states statically, using weakest precondition reasoning. The comparison of the two program states illustrates the source of the error; a change to the program that reconciles the two states fixes the bug. Weakest precondition reasoning allows for a quite detailed characterization of the states, but it also requires to start with a strong postcondition (a full functional specification), whereas methods based mostly on dynamic analysis—such as code-based fixing—provide approximate yet useful characterization even with very weak formal specifications.

### 5.3 Our Previous Work

As part of the AUTOFIX joint project between ETH Zurich and Saarland University, we developed the tools Pachika and AutoFix-E. Pachika [5] automatically builds finite-state behavioral models from a set of passing and failing test cases of a Java program. Pachika also generates fix candidates by modifying the model of failing runs in a way which makes it compatible with the model of passing runs. The modifications can insert new transitions or delete existing transitions to change the behavior of the failing model; the changes in the model are then propagated back to the Java implementation.

AutoFix-E [21] implements the first automatic program fixing tool for Eiffel, based on model-based techniques. AutoFix-E uses argumentless boolean queries to abstract the object space, hence it works best for classes with a detailed interface. Code-based techniques improve on model-based ones by locating faults based on both dynamic and static analysis techniques.

## 6 Future Work

Future work includes the following aspects:

- All our experiments with automated fixing involve automatically generated test cases, but state-of-the-art random testing is not applicable to every type of program; for instance, applications involving input through files or an interactive graphical interface are arduous to test automatically. We plan to experiment our techniques for automated fixing on new types of software with manually written test cases.

- A non-negligible portion of the bugs found in EiffelBase, likely representative of much of software written in Eiffel, are due to incorrect *contracts* rather than implementations. We will try to flip over our approach to program fixing and fix contracts when the implementation is correct. This effort is tightly related to our other work on contract inference [20].

- Applying program fixing techniques to languages without contracts requires to consider other types of faults to fix, including exceptions and I/O errors. We will extend AutoFix-E$_2$ to handle these types of faults.

- While the majority of bugs can be fixed with a small patch, there exist conspicuous errors that require significant changes to the code. We plan to apply more ambitious code synthesis techniques to the problem of building a fix once the "cause" of a fault is known.

- AutoFix-E2 is part of the Eve verification environment [8]. As part of the ongoing improvements of Eve, we will ameliorate the user interface of AutoFix-E2 and its integration with Eve's other verification aides.

# 7 Conclusion

This paper introduced code-based automated program fixing, a novel approach to generate automatically corrections of errors in software equipped with contracts. Preliminary experiments with the supporting tool AutoFix-E2 demonstrate that code-based techniques extend the applicability of automated program fixing to more faults in classes beyond well-designed data structure implementations.

**Availability.** The AutoFix-E2 source code, and all data and results cited in this article, are available at:

> http://se.inf.ethz.ch/research/autofix/

**Acknowledgments.** This work is partially funded by Hasler-Stiftung Grant no. 2327 "AutoFix—Programs that fix themselves". The facilities provided by the Swiss National Supercomputing Centre (CSCS) ran AutoTest to generate some of the test cases used in the experiments. The authors thank Andreas Zeller for the ongoing collaboration on automated program fixing. Bernhard Brodowsky, Severin Heiniger, and Stefan Heule implemented the text document library.